\documentclass[aps,prx,twocolumn,preprintnumbers,amsmath,amssymb,longbibliography]{revtex4-2}
\usepackage{dcolumn}
\usepackage{bm}
\usepackage{graphicx}
\usepackage{amsmath,amsfonts,amssymb,amsthm,amstext,amsbsy,amsopn,amscd,amsxtra}
\usepackage{dsfont}
\usepackage{physics} % for \ket{}
\usepackage{upgreek} % for e.g. \upmu (straight greek letter)
\usepackage[colorlinks=true,allcolors=blue]{hyperref}
\usepackage{xcolor}
\usepackage{url}
\usepackage{siunitx}
\usepackage{multirow}
\def\epsilon{\varepsilon}
\def\F{{\cal F}}

\def\degree{^\mathrm{o}}

\everymath{\displaystyle}

\begin{document}

\title{Quantum state control of a Bose-Einstein condensate in an optical lattice}

\author{N. Dupont$^1$, G. Chatelain$^1$, L. Gabardos$^1$, M. Arnal$^1$, J. Billy$^1$, B. Peaudecerf$^1$, D. Sugny$^2$, D. Gu\'ery-Odelin$^1$}
\email{dgo@irsamc.ups-tlse.fr}
\affiliation{
$^1$ Laboratoire Collisions Agr\'egats R\'eactivit\'e, UMR 5589, FERMI, UT3, Universit\'e de Toulouse, CNRS,\\
118 Route de Narbonne, 31062 Toulouse CEDEX 09, France \\
$^2$ Laboratoire Interdisciplinaire Carnot de Bourgogne, UMR 6303,\\
9 Avenue A. Savary, BP 47 870, F-21078 Dijon Cedex, France
}

\date{\today}

\begin{abstract}
We report on the efficient design of quantum optimal control protocols to manipulate the motional states of an atomic Bose-Einstein condensate (BEC) in a one-dimensional optical lattice. Our protocols operate on the momentum comb associated with the lattice. In contrast to previous works also dealing with control in discrete and large Hilbert spaces, our control schemes allow us to reach a wide variety of targets by varying a \emph{single} parameter, the lattice position. With this technique, we experimentally demonstrate a precise, robust and versatile control: we optimize the transfer of the BEC to a single or multiple quantized momentum states with full control on the relative phase between the different momentum components. This also allows us to prepare the BEC in a given eigenstate of the lattice band structure, or superposition thereof.
\end{abstract}

\maketitle

\section{Introduction}

Quantum simulation amounts to using a tunable quantum system to simulate another less controllable one, and is particularly suited for studying model dynamics in large Hilbert spaces~\cite{PRXQuantum.2.017003}. 
This field has been growing very fast in the past years as an alternative to the all-purpose quantum computer and quantum simulations are currently under study on a wide variety 
of quantum platforms (see~\cite{PRXQuantum.2.017003,RMPNori,Cirac2012,Arguello2019} and references therein).

Key requirements for quantum simulation are the ability to generate the desired Hamiltonian, to reliably prepare the initial quantum state of interest, and to measure the subsequent time-evolved state. 
% note: the serial comma is recommended by Phys Rev
Many innovative techniques have been developed to produce synthetic Hamiltonians, using, e.g., properly designed coupling
 between external fields and atoms~\cite{RMPNori,dalibard11,dalibard19} or periodic driving of the system's parameters~\cite{FloquetRMP,Nathan}.
The time evolution can either be probed continuously, or investigated at fixed time intervals in the case of a digitized time evolution \cite{Lanyon57} or to study discretized dynamics \cite{arnal2020,timecrystal}.

Satisfying the requirement of the reliable preparation of a desired initial state often calls for manipulating a dynamical system by means of external time-dependent parameters. In practice it amounts to steering that system from a state easily provided by an experimental setup to the desired one in a minimum time and with a very high fidelity, while satisfying experimental constraints and limitations. Many quantum control methods such as shortcuts to adiabaticity~\cite{reviewSTA}, composite pulses~\cite{genov2014}, machine learning approaches~\cite{wigley2016} or protocols based on optimal control theory with a minimal use of resources \cite{liberzon,bonnardbook,pont,glaserreview,boscain2021} have been developed and even combined to reach this goal in a variety of contexts.

In particular, optimal control theory, which has its roots in engineering~\cite{bonnardbook}, has been transposed with success to quantum systems first in the context of physical chemistry~\cite{brifreview,RMP19}, then applied in Nuclear Magnetic Resonance for the control of spin dynamics~\cite{nielsen2010,grape,lapert:2010}, and nowadays in quantum technologies~\cite{glaserreview}. For quantum systems with complex dynamics, efficient iterative algorithms have been developed to solve optimal control problems~\cite{grape,reichkrotov,gross}. 
This has recently led to experimental realizations of optimal control protocols aimed at engineering quantum states or operations in various platforms, such as NV centers~\cite{waldherr2014,nobauer2015}, photonic states in a cavity~\cite{heeres2017}, internal states of atoms in low-excitation states~\cite{jessen2015,saywell2020} or in a Rydberg manifold~\cite{HarochePRX2020}.

\medskip

Atomic Bose-Einstein condensates (BECs), in which all atoms occupy a macroscopic wave function, lend themselves naturally to the use of optimal control techniques.
Theoretical studies have investigated how to manipulate BECs  via a modulation of the magnetic confinement potential in a one-~\cite{borzi2007,jager2014,sorensen2018,hocker2016} but also three-dimensional case~\cite{mennemann2015}, and on atom chips~\cite{corgier2018,amri2019,chen2011,zhang2016}, outperforming simpler control protocols.  Experimental evidence of the efficiency of optimal control laws has been provided for the manipulation of vibrational states in a single trap~\cite{bucker2013,frank2016}, in particular for interferometry applications~\cite{frank2014}. Other investigations extended to the optimal driving of strongly interacting atoms, including through a phase transition~\cite{doria2011,frank2016}.

The control of a BEC in optical lattices is of particular interest for quantum simulation~\cite{groosLattices2017}. Simple momentum state superpositions can be generated~\cite{potting2001} in order to realize a shaken-lattice interferometer~\cite{weidner2017,weidner2018a,weidner2018}. Optimal control at quantum speed limit in a two-level quantum system realized in a lattice was demonstrated in~\cite{bason2011}. Efficient control targeting specifically the eigenstates of optical lattices has been achieved in~\cite{zhou2018}, but required combined optimized manipulations of both the lattice position and amplitude.

\medskip 

In this study, we use quantum optimal control (QOC) protocols to shape the momentum distribution of a BEC in a one-dimensional optical lattice. The Hilbert space has a large unbound size provided by a momentum comb of fixed spacing. In contrast with previous studies \cite{bucker2013,frank2014,jessen2015,heeres2017,HarochePRX2020}, we drive the system through this large space by continuously varying a \emph{single} parameter, the lattice position.
With this simple but sufficient scheme, we experimentally achieve a robust and versatile control of the BEC wave function, and prepare arbitrary states, in a short time window (typically of the order of \SI{100}{\micro\second}) compared to commonly used adiabatic protocols. This allows us to reach a wide variety of states for which there is no known analytical solution of the control problem. We demonstrate the transfer of the BEC to single and multiple quantized momentum states, with the ability to control the relative phases between the different momentum components. As a further application, we prepare the BEC in arbitrary superpositions of Bloch eigenstates of the lattice. The implementation of the computed optimal control systematically leads to an excellent agreement between the numerical target and the experimental result.

\section{Experimental setup and algorithm}

\subsection{BEC Experimental setup}

We perform our experiments in a hybrid trap~\cite{fortun2016} in which we obtain pure rubidium-87 BECs of typically $2\cdot 10^5$ atoms in the lowest hyperfine state $|F=1, m_F = -1\rangle$. These BECs are loaded in a one-dimensional optical lattice produced by two laser beams of wavelength $\lambda=\SI{1064}{\nano\meter}$ counterpropagating along the $x$-axis. They are superimposed on the optical dipole beams and the quadrupolar magnetic field of the hybrid trap. Along the optical lattice axis, the atoms experience the potential

\begin{equation}
	U(x,t) = -\frac{s}{2}E_\mathrm{L}\cos\left(k_\mathrm{L} x + \varphi(t)\right) + U_\mathrm{hyb}(x),
	\label{eq:potential}
\end{equation}
where $k_\mathrm{L} \equiv 2\pi/d=2\pi/(\lambda/2)$ and $E_\mathrm{L} = \hbar^2k_\mathrm{L}^2/(2m)$ (with $m$ the mass of an atom and $\hbar$ the reduced Planck constant) are respectively the wavevector and the energy associated with the lattice ($E_\mathrm{L}=4E_\mathrm{R}$, with $E_\mathrm{R}$ the recoil energy). The harmonic potential of the hybrid trap $U_\mathrm{hyb}(x)$ has an angular frequency $\omega_x=2\pi \times 50$ Hz. 
The dimensionless depth of the lattice $s$ is independently and precisely calibrated~\cite{CabreraCalibration} for each experiment presented here. 

We adiabatically load the atoms in the ground state of the lattice potential, i.e. in a Bloch wave of spatial period $d$ \cite{chatelain2020}. As a result, the momentum distribution is made up of equally spaced peaks separated by $\hbar k_\mathrm{L}$. Experimentally, we image this momentum distribution after a sufficiently long ballistic expansion following a sudden switch off of the confining potentials. The final spatial density $n(\mathbf{r})$ then reproduces the initial momentum density $\tilde{n}(\mathbf{p})$ up to a scaling factor: $n(\mathbf{r},t=t_\mathrm{TOF})=\tilde{n}(\mathbf{p}=m\mathbf{r}/t_\mathrm{TOF},t=0)$, with $t_\mathrm{TOF}$ the time-of-flight duration.

The phase $\varphi(t)$ of the lattice potential is our control parameter. Varying $\varphi$ as a function of time amounts to moving the lattice position along the $x$-axis. This parameter is set by the relative phase between the two phase-locked acousto-optic modulators controlling the lattice beams.  In the following section, we explain how to engineer $\varphi(t)$ to tailor at will the momentum distribution. As the phase is varied, we only apply global transformations to the lattice: as a consequence the quasi-momentum $q$ in the laboratory reference frame remains equal to its value for the initially prepared ground state of the lattice, $q=0$. Therefore, in that reference frame, we engineer arbitrary momentum superposition states of the form
\begin{equation}
|\Psi \rangle=\sum_{\ell\in \mathds{Z}} c_\ell \ket{\chi_{\ell}},
\end{equation}
where the vector $\ket{\chi_\alpha}$ is the eigenstate of the momentum operator with eigenvalue $\alpha \, \hbar k_\mathrm{L}$ (and $\chi_\alpha(x)=e^{i\alpha k_\mathrm{L}x}/\sqrt{2\pi}$).

\subsection{Control algorithm}
\label{sec:algo}

Our objective is to tailor the time variation of the control field $\varphi(t)$ over a fixed time interval $[0,t_\mathrm{f}]$, in order to assign a desired value to the $c_\ell$ coefficients. For this purpose, we use a standard first-order gradient algorithm~\cite{bryson,gross,grape}. As the quasi-momentum $q=0$ is fixed in the laboratory reference frame, we recast the Schr\"odinger equation involving \emph{only} the optical lattice potential in the following matrix form:
\begin{equation}
i\dot{C}=\mathcal{M}(\varphi(t))C,
\end{equation}
where the state of the system is denoted by the vector $C$ with coordinates $\{c_\ell\}$ which satisfy
\begin{equation}
i\dot{c}_{\ell}=\ell^2c_{\ell}-\frac{s}{4}(e^{i\varphi(t)}c_{\ell-1}+e^{-i\varphi(t)}c_{\ell+1}),
\end{equation}
with a rescaling of time: $t\rightarrow E_\mathrm{L}t/\hbar$.
The initial condition is given by the vector associated to the lowest Bloch state at $q=0$ for the lattice depth $s$.
The optimal control problem is defined with respect to a figure of merit $\mathbb{F}(C(t_\mathrm{f}),C^\dagger(t_\mathrm{f}))$ to be maximized at the fixed final time $t_\mathrm{f}$ of the control process. The choice of the function $\mathbb{F}$  depends on the objective of the control (see below). We do not put any other constraint on the control field $\varphi(t)$. The optimal solution is formulated from the Pontryagin Hamiltonian~\cite{boscain2021,liberzon}
\begin{equation}
H_\mathrm{P}=\Re [\langle D| \dot{C}\rangle]=\Im[\langle D|\mathcal{M}|C\rangle],
\end{equation}
where $\Re[\cdot]$ and $\Im[\cdot]$ denote respectively the real and the imaginary parts of a complex number. We introduce $D$, the adjoint vector, solution of
\begin{equation}\label{eqadj}
i\dot{D}=\mathcal{M}(\varphi(t))D,
\end{equation}
with the final condition
\begin{equation}
\label{eq:finalD}
D(t_\mathrm{f})=\frac{\partial \mathbb{F}(C(t_\mathrm{f}),C^\dagger(t_\mathrm{f}))}{\partial C^\dagger(t_\mathrm{f})}.
\end{equation}
The correction $\delta\varphi$ to first order of the control field is proportional to the derivative of $H_\mathrm{P}$ with respect to $\varphi(t)$
\begin{equation}
\delta H_\mathrm{P} (t)=\frac{\partial H_\mathrm{P}}{\partial \varphi(t)}=\Im \left[\langle D|\frac{\partial \mathcal{M}}{\partial \varphi(t)}|C\rangle\right].
\end{equation}
We thus consider the following gradient algorithm:
\begin{enumerate}
\item Choose a guess field $\varphi(t)=\varphi_0(t)$.
\item Propagate forward the state $C(t)$ up to the final state $C(t_\mathrm{f})$.
\item Propagate backward the adjoint state of the system from Eq.~\eqref{eqadj}, with the final condition given by Eq.~\eqref{eq:finalD}.
\item Compute the correction $\delta \varphi(t)$ to the control field, $\delta \varphi(t)=\epsilon \delta H_\mathrm{P}(t)$, where $\epsilon$ is a small positive parameter.
\item Define the new control field $\varphi(t)\mapsto \varphi(t)+\delta \varphi(t)$.
\item Repeat from step 2 until the desired value of the figure of merit $\mathbb{F}$ is reached.
\end{enumerate}
In the numerical simulations, $\epsilon$ is determined from a line search method~\cite{bryson}. A constant phase adapted to the target state is generally a good starting control field for the algorithm. For the experimental implementation under study, we have verified that the accuracy of a first-order algorithm is sufficient, and second-order corrections are therefore not necessary~\cite{fouquieres2011,dalgaard2020}. 

\subsection{Implementation}

In practice the optimal control field $\varphi(t)$ manipulated by the algorithm is a piecewise constant function. The control sequence is divided into a large number of time steps $\Delta t$ (each of the order of a few hundreds of ns). The relatively smooth optimal pulses derived from the algorithm justifies this approximation. 
Its successful implementation in the experiment imposes stringent technical requirements, such as the ability to vary the phase in an arbitrary fashion with an accuracy of less than a percent and on a frequency scale much higher than the characteristic frequencies of the atomic dynamics ($\sim 10$ kHz)
\footnote{In this work this is achieved through the use of DDS synthesizers (\textsc{Keysight 33612a}), computer-controlled and synchronized on a same clock signal. This typically give us the ability to sample modulation ramps with frequencies up to \SI{3}{\mega\hertz}, in steps as low as \SI{2}{\nano\second}. These modulation ramps are then applied to the RF drive of the acousto-optic modulator controlling the phase $\varphi(t)/2$ of the lattice light.
}.

Finally, the choice of the time duration $t_\mathrm{f}$ of the phase control is a compromise between two opposite requirements. It should be larger than the natural time scale of evolution for the atoms in the lattice since the algorithm takes advantage of the dynamics, but not too long to mitigate the deleterious effects of the accumulation of small errors in the control, due to experimental fluctuations or imperfections. In practice, we choose a control time $t_\mathrm{f}\sim 1.5\,T_0$, where $T_0$ is the time associated with the energy difference between the two lowest eigenstates at $q=0$ ($T_0\simeq$\,\SI{70}{\micro\second} for $s\simeq 5$). This experimentally leads to the highest fidelities to the desired targets.

\section{Control of momentum state populations}
\label{secIII}

In a first set of experiments, we investigate the control of the populations $p_\ell=|c_\ell|^2$ in the desired momentum orders ($\textstyle \sum_{\ell} p_\ell=1$). To reach a specific target with populations $\lbrace p_{\mathrm{t}, \ell} \rbrace$, we run the previously described algorithm with the figure of merit
\begin{align}
\label{eq:f-o-mp}
\mathbb{F} &= 1-\frac{1}{2}\sum_\ell(|c_\ell|^2-p_{\mathrm{t}, \ell})^2\\
&= 1-\frac{1}{2}\sum_\ell(C^\dagger O_\ell C-p_{\mathrm{t}, \ell})^2, \notag
\end{align}
where $O_\ell$ is the projection operator onto the state $\ell$~\cite{jozsa1994}. Using this definition, and the algorithm described in Sec.~\ref{sec:algo}, we present optimal control ramps leading first to the preparation of a single momentum state (Sec.~\ref{sec:popssingle}), and then to arbitrary-weight superpositions of two or more momentum components (Sec.~\ref{sec:popsmany}).
%. Note that other definitions of the quantum fidelity of mixed states could be used with equivalent results~
\subsection{Populating a single momentum state}
\label{sec:popssingle}

%Aiming at a first type of targets, 
We first optimize the control field $\varphi(t)$ to populate a single momentum state $\ket{\chi_{n}}$. In doing so, we somehow ``erase" in momentum space the information on the periodicity of the wavefunction. We compare the experimentally obtained momentum states to the targeted ones by computing the average fidelity
\begin{equation}
\label{eq:fidelityp}
\F(p_\mathrm{exp},p_\mathrm{t}) =\left[ \sum_{\ell} \sqrt{p_{\mathrm{exp},\ell}}\sqrt{p_{\mathrm{t},\ell}} \right]^2,
\end{equation}
where the $p_{\mathrm{exp},\ell}= |c_{\mathrm{exp},\ell}|^2$ correspond to the populations of the different momentum components of the experimental distribution $p_\mathrm{exp}$, obtained by averaging over 10 experimental realizations. The ideal targeted momentum distribution $p_\mathrm{t}$ is here given by $p_\mathrm{t,\ell}=\delta_{\ell,n}$. We compare the experimental fidelity to the theoretical value $\F_\mathrm{num}$ obtained from numerical simulations using the same control field $\varphi(t)$. We use this fidelity to assess our results rather than the figure of merit of Eq.~\eqref{eq:f-o-mp} as it is closer to the usually defined fidelity in quantum physics (the modulus squared of the states overlap). The chosen figure of merit for the algorithm leads however to a simpler gradient computation, and both quantities should reach the value 1 for an optimal control solution.

\begin{figure}[ht]
	\begin{center}
		\includegraphics[scale=1]{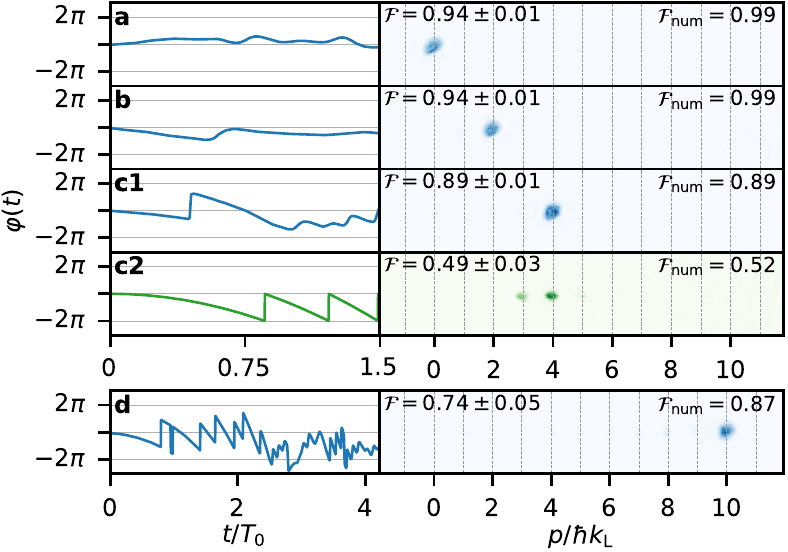}
		\caption{\textbf{Single momentum states.} (\textbf{a-d}) Left: Control fields $\varphi(t)$ for the preparation of single momentum states, respectively $0,2,4$ and $10 \, \hbar k_\mathrm{L}$. Time is given in units of $T_0$ (see text). (\textbf{a}), (\textbf{b}), (\textbf{c1}) and (\textbf{d}) (blue) are obtained by QOC and (\textbf{c2}) (green) is the quadratic phase evolution of a uniformly accelerated lattice reaching velocity $4 \, h / md$ in $1.5 \, T_0$ (see text). Right: corresponding absorption images of diffraction orders. Displayed fidelities are computed with respect to the ideal single momentum target states, with $\F$ the average fidelities of 10 experimental realizations (errors are statistical and correspond to one standard deviation) and $\F_\mathrm{num}$ the expected fidelities from numerically evolving the state in the lattice potential with the corresponding $\varphi(t)$. The calibrated dimensionless lattice depths are $s_\mathrm{\textbf{a}} = 5.1 \pm 0.2$, $s_\mathrm{\textbf{b}} = 5.2 \pm 0.2$, $s_\mathrm{\textbf{c1}} = 5.2 \pm 0.2$, $s_\mathrm{\textbf{c2}} = 5.0 \pm 0.1 $ and $s_\mathrm{\textbf{d}} = 5.1 \pm 0.2$.}
		\label{fig:monomes}
	\end{center}
\end{figure}

In the experiments shown in Fig.~\ref{fig:monomes}, we choose a lattice depth $s\simeq5$ and populate various positive single momentum states up to the momentum state $p=10 \, \hbar k_\mathrm{L}$.  The measured experimental fidelities are very close to the numerical ones, and  both show a similar trend, with slowly decreasing values as we go from low momentum targets to high momentum targets. At the same time, the control phase profile gets more complex, as seen in Fig.~\ref{fig:monomes}\,(\textbf{d}). To reach high single momentum states, such as $p=10 \, \hbar k_\mathrm{L}$ (Fig.~\ref{fig:monomes}\,(\textbf{d})) we had to increase the time of the control ramp to ensure the convergence of the algorithm. We recover here a standard result of optimal control, namely that the reachable set increases for larger control times. It is worth noticing that the QOC algorithm allows us to reach momentum values that are much higher than the ones accessible from the classical free dynamics \cite{chatelain2020}, for which $p_\mathrm{max}^2 \sim 2m\,sE_\mathrm{L}$ i.e. $p\mathrm{max}\sim\sqrt{s}\,\hbar k_\mathrm{L}\simeq 2.2\,\hbar k_\mathrm{L}$. In using QOC to populate with high fidelity a single diffraction order $n$ of the lattice, we effectively realize a blazed grating for matter waves, with the algorithm designing a constant phase gradient $\Delta \phi(x)=n\,k_\mathrm{L} x=2\pi n \, x/d$ across the lattice sites. 

%Another method to populate high order momentum states is to accelerate the lattice.
We can compare our QOC method to the more standard protocol of accelerating the lattice to impart momentum to the atoms.
We illustrate this comparison in the case of $n=4$ in Fig.~\ref{fig:monomes}\,(\textbf{c2}) : we uniformly accelerate the lattice up to a velocity of $4\,h/(md)$ for the same amount of time $t_\mathrm{f}$ and lattice depth $s$ used for the QOC experiment. After the acceleration, the fidelity to the target state, i.e. the single momentum state $p=4\,\hbar k_\mathrm{L}$, is much lower than the one obtained using the QOC method. This is clearly visible on the experimental absorption image, with a noticeable population on the $p=3\,\hbar k_\mathrm{L}$ momentum peak for the acceleration method \cite{Battesti2004}.
One could argue that the fidelity to the target state using the acceleration method can be increased by working adiabatically and performing Bloch oscillations. Numerical simulations show however that reaching a fidelity $>$ 90\% to a given single momentum state requires low lattice depths $s<1$ and that the adiabaticity condition then leads to a generally much longer control time $t_\mathrm{f}$. The control time will also grow linearly with the momentum of the targeted state, reaching $t_\mathrm{f}=1.7\,$ms for the $p=4\,\hbar k_\mathrm{L}$ state considered here. 
Such long control times also mean that the atomic wavepacket moves much further away from the center of the hybrid trap during its acceleration, with deleterious effects. We had in fact to reduce the axial confinement from \SI{50}{\hertz} to \SI{4}{\hertz} when using the acceleration method to achieve the result of Fig.~\ref{fig:monomes}\,(\textbf{c2}).
In Tab.~\ref{table1}, we summarize the results of such a comparison 
between optimal control and Bloch acceleration
for several momentum orders. 
We conclude that the QOC method is both a fast and accurate procedure to populate single momentum states with a high fidelity.

\begin{table}[ht]
\begin{tabular}{|c|c|cccc|}
\hline
\multicolumn{2}{|c|}{Target $n$} & 2    & 4    & 8 &  10   \\ \hline
\multicolumn{2}{|c|}{Fidelity $\mathcal{F}$}      & $0.94{\scriptstyle\pm 0.01}$ & $0.89{\scriptstyle\pm 0.01}$ & $0.76{\scriptstyle\pm 0.04}$ & $0.74{\scriptstyle\pm 0.05}$ \\ \hline
\multirow{2}{*}
{QOC}         & $s(\pm 0.2)$      & 5.2  & 5.2  & 5.1 & 5.1  \\
                            & $t_\mathrm{f}$ ($ \si{\micro\second}$)     & 91.7 & 91.7 & 260 & 260  \\ \hline
\multirow{2}{*}
{Acc.}       & $s$      & 0.75 & 1.1 & 2.1 & 2.3 \\
                            & $t_\mathrm{f}$ ($\si{\micro\second}$)     & $1.7 \cdot 10^3$  & $1.7 \cdot 10^3$  & $1.2 \cdot 10^3$ & $1.4 \cdot 10^3$ \\ \hline
\end{tabular}
\caption{Comparison of the control times $t_\mathrm{f}$ required to reach a given single momentum target ($p=n\,\hbar k_\mathrm{L}$) with the same fidelity using either our Quantum Optimal Control (QOC) protocol or using the quasi-adiabatic Bloch oscillation scheme with a uniformly accelerated lattice (Acc.). We indicate the lattice depth $s$ at which the experiments were performed (for the QOC case), or which would be required (for the accelerated case).}
\label{table1}
\end{table}

The comparison between the control fields $\varphi(t)$ in both methods sheds light on the way the optimal phase is designed: the folded quadratic growth pattern observed in the control phase for the acceleration method can also be found in the QOC phases, for instance in Fig.~\ref{fig:monomes}\,(\textbf{c1}) (and also in (\textbf{d})). In this case, the optimal control field $\varphi(t)$ can be interpreted in terms of a first acceleration stage ($t/T_0<0.75$) towards the targeted momentum state, and a second ``correction" stage to reduce the population in unwanted momentum states.

The QOC solution turns out to be quite robust. Indeed, we get small error bars on the measured average fidelity over 10 realizations, despite the possible shot-to-shot atom number fluctuations or lattice depth fluctuations (there is no active stabilization of the lattice beam intensity), the atom-atom interactions that affect the BEC wave function, and the presence of the external confinement $U_\mathrm{hyb}(x)$ superimposed on the lattice which is not taken into account in the algorithm. 

\subsection{Populating an arbitrary number of momentum states}
\label{sec:popsmany}
\begin{figure}[ht]
	\begin{center}
		\includegraphics[scale=1]{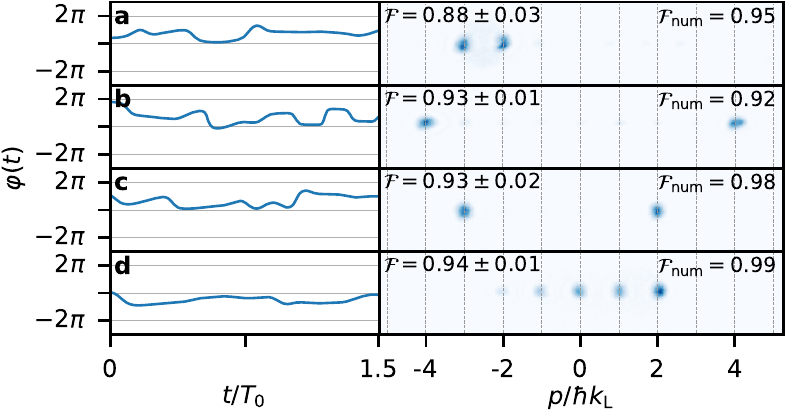}
		\caption{\textbf{Arbitrarily populated momentum states.} (\textbf{a-c}) Equiprobable superpositions of respectively $\left( -3, -2 \right)\hbar k_\mathrm{L}$, $\left( -4, 4 \right)\hbar k_\mathrm{L}$ and $\left( -3, 2 \right) \hbar k_\mathrm{L}$ momentum states. (\textbf{d}) Superposition of the momenta $\left( -2, -1, 0, 1, 2\right) \hbar k_\mathrm{L}$ with populations $p_\ell = \left\lbrace 1, 2, 3, 4, 5 \right\rbrace/15$. Left: QOC computed control fields $\varphi(t)$. Time is given in units of $T_0$ (see text). Right: corresponding absorption images of the momentum distribution. Displayed fidelities are computed with respect to the ideal target states, with $\F$ the fidelities from an average over 10 experimental realizations (errors are statistical and correspond to one standard deviation) and $\F_\mathrm{num}$ the expected fidelities from numerically evolving the state in the lattice potential with the corresponding $\varphi(t)$. The calibrated dimensionless lattice depths are $s_\mathrm{\textbf{a}} = 4.6 \pm 0.2$, $s_\mathrm{\textbf{b}} = 5.0 \pm 0.1$, $s_\mathrm{\textbf{c}} = 4.6 \pm 0.2$ and $s_\mathrm{\textbf{d}} = 5.1 \pm 0.1 $.}
		\label{fig:multinomes}
	\end{center}
\end{figure}

In a second set of experiments, we shape the phase $\varphi(t)$ to populate an arbitrary number of momentum states with the desired probabilities. First, we realize equiprobable superpositions of two momentum states, varying their relative momentum: we show the case of neighbouring momentum states (Fig.~\ref{fig:multinomes}\,(\textbf{a})), opposite momentum states (Fig.~\ref{fig:multinomes}\,(\textbf{b})) and an arbitrary pair of momentum states (Fig.~\ref{fig:multinomes}\,(\textbf{c})) for which an earlier simple method (with a constant phase and a specific control time) gave decent results in specific cases~\cite{chatelain2020}. A closer look at the absorption image obtained for two neighbouring momentum states reveals the presence of a diffuse scattering halo, which results from elastic collisions occuring during the time-of-flight (see Ref.~\cite{chatelain2020} and references therein). We also demonstrate the superposition of a high number of momentum states (five in Fig.~\ref{fig:multinomes}\,(\textbf{d})) with arbitrary weights. In each case shown, we achieve good experimental fidelities to the ideal target (larger than $88\%$), just slightly below the corresponding numerical fidelities. The QOC robustness and versatility also allowed us to record all $2^7=128$ combinations of equal-weight superpositions of momenta between $p=-3\,\hbar k_\mathrm{L}$ and $p=3\,\hbar k_\mathrm{L}$, which can then e.g. be stacked together to spell out words (see Appendix \ref{appen:printer}).

\section{Full state control of momentum superpositions}

In Sec.~\ref{secIII}, the control field $\varphi(t)$ was optimized toward a target defined solely in terms of the populations ${p_\ell}$ of the momentum peaks regardless of their relative phases. In this section, we improve the degree of control by targeting both the populations of and the relative phases between different momentum components. To fulfill these requirements, we run our optimization algorithm with a figure of merit sensitive to phase differences, the standard quantum fidelity: 
\begin{equation}
\mathbb{F}=\mathcal{F}_Q=|\langle C(t_\mathrm{f})|C_\mathrm{t}\rangle| ^2,
\end{equation} 
with $C_\mathrm{t}$ the target vector.

In the following, we first show that we can prepare and identify a superposition of two momentum components with an arbitrary relative phase (section \ref{sec:phase}). Then we move to the preparation of more complex superpositions of momenta, such as specific eigenstates of the lattice potential, and superpositions thereof (Sec.~\ref{sec:eigen}).

\subsection{Controlling the phase of a momentum superposition}
\label{sec:phase}

\begin{figure*}[ht]
	\begin{center}
		\includegraphics[scale=1]{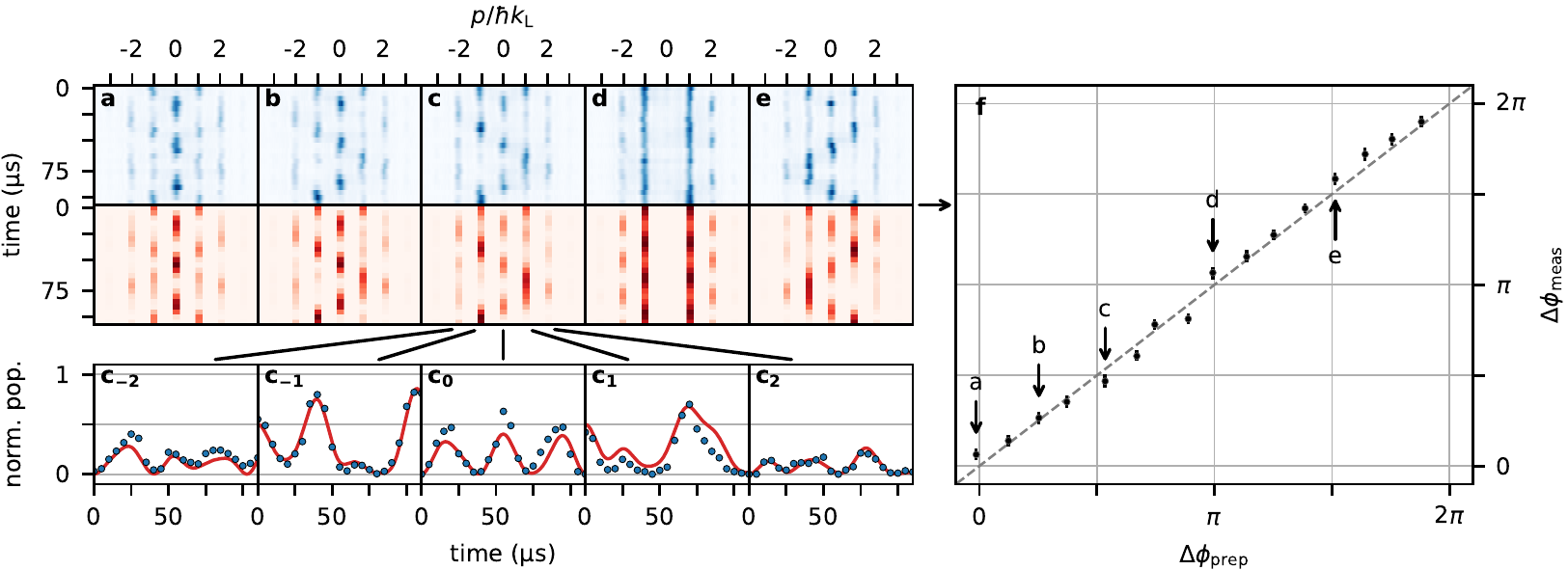}
		\caption{\textbf{Control and measurement of the phase between momentum components.} (\textbf{a-e}) Top: stacks of integrated experimental images (blue) showing the evolution of the momentum distribution during a \SI{110}{\micro\second} holding time in a static lattice after applying a control field $\varphi(t)$ preparing the momentum superposition $\ket{\Psi_{\Delta\phi}}$ of momentum components $\left( -1, 1 \right)\hbar k_\mathrm{L}$ (see text) with an expected relative phase $\Delta \phi_\mathrm{prep} = 3\degree$, $46\degree$, $96\degree$, $184\degree$ and $276\degree$. Bottom: numerical propagation (red) in a static lattice of the same momentum superposition with a relative phase measured by least-squares fitting of the experimental data (see text), yielding respectively $\Delta \phi_\mathrm{meas} = \left(11 \pm 6\right)\degree$, $\left(48 \pm 7\right)\degree$, $\left(84 \pm 6\right)\degree$, $\left(192 \pm 7\right)\degree$ and $\left(285 \pm 6\right)\degree$. (\textbf{c$_{-\mathbf{2}}$-c$_\mathbf{2}$}) Detail of the evolution of momentum populations in (\textbf{c}), with panel (\textbf{c$_\emph{i}$}) featuring the $i$-th momentum component, and showing the experimental data (blue dots) and numerical propagation of the superposition $\ket{\Psi_{\Delta\phi_\mathrm{meas}}}$ with relative phase $\Delta\phi_\mathrm{meas}$ determined by least-square fitting (continuous red line). (\textbf{f}) Measured relative phase $\Delta\phi_\mathrm{meas}$ as a function of QOC prepared relative phase $\Delta\phi_\mathrm{prep}$ for data (\textbf{a-e}) and more. All data shown were obtained for a calibrated lattice depth $s\simeq5$. The error bars represent the 95\% confidence interval for the value of $\Delta\phi_\mathrm{meas}$ deduced from the likelihood function. The grey dotted line is of slope one.
		%({\color{teal} comment error bars})
		}
		\label{fig:dphi}
	\end{center}
\end{figure*}

To demonstrate control over the relative phase of momentum components, we focus here on a simple momentum superposition of the form:

\begin{equation}
\ket{\Psi_{\Delta\phi}}=\frac{1}{\sqrt{2}}(\ket{\chi_{1}}+e^{i\Delta\phi}\ket{\chi_{-1}}).
\end{equation}

For several values of the relative phase $\Delta\phi_j=j\times\pi/8,\:j\in \{0,\cdots,15\}$, we find an optimal control ramp that prepares the corresponding superposition $\ket{\Psi_{\Delta\phi_j}}$. To identify the relative phase of the prepared superposition we use the subsequent evolution of the momentum distribution in the static lattice: just after the preparation ramp, the lattice phase is set back to its initial value, $\varphi(t>t_\mathrm{f})=0$, and we measure the evolution of the momentum distribution over an extra \SI{110}{\micro\second}. This evolution depends strongly on the effective phase of the superposition, as illustrated in Fig.~\ref{fig:dphi}\,(\textbf{a-e}). All experiments are performed in a lattice of depth $s\simeq5$ \cite{CabreraCalibration}.

A precise determination of the phase of the superposition is achieved by fitting the time-evolution of the ideal state $\ket{\Psi_{\Delta\phi}}$ to the measured evolution of the momentum orders, as shown in Figs.~\ref{fig:dphi}\,(\textbf{$\mathrm{\mathbf{c_{-2}-c_{2}}}$}). A least-square fitting with $\Delta\phi$ as free parameter yields the measured phase $\Delta\phi_\mathrm{meas}$. In Fig.~\ref{fig:dphi}\,(\textbf{f}), we compare this measured phase to the prepared phase $\Delta\phi_\mathrm{prep}$, which is the phase of the superposition as given by the optimal control solution (which may differ slightly from the target $\Delta\phi_j=j\times\pi/8$), for all the values $\Delta\phi_j$. The error bars represent the 95\% confidence interval determined from the likelihood function \cite{james2006}. The good agreement between the phase expected from the optimal control law $\Delta\phi_\mathrm{prep}$ and the measured result  $\Delta\phi_\mathrm{meas}$ demonstrates our ability to engineer the phase of momentum superpositions reliably.
%From the residuals between expected and measured phase in Fig.~\ref{fig:dphi}\textbf{f}, we find that the experimental implementation of the control procedure yields the desired phase of the superposition within a $8^{\circ}$ standard deviation.
Examples of three-momenta superpositions with different relative phases are also provided in Appendix \ref{appen:three}.

\subsection{Preparing lattice eigenstates}
\label{sec:eigen}

To further illustrate our control on the relative phase in momentum superpositions, we use the very same optimal control algorithm to prepare eigenstates of the lattice potential. For a given lattice quasi-momentum $q\equiv\tilde{q}k_\mathrm{L}$, the $n^{\mathrm{th}}$ Bloch function reads:
\begin{equation}
\ket{\psi_{n,q}}=\sum_\ell c_\ell^{(n,q)} \ket{\chi_{\ell+\tilde{q}}},
\label{eq:bloch}
\end{equation}
where the coefficients $c_\ell^{(n,q)}$ are
solutions of the stationary Schr\"odinger equation
\begin{equation}
E_{n,q}{c}_{\ell}=(\ell+\tilde{q})^2c_{\ell}-\frac{s}{4}(c_{\ell-1}+c_{\ell+1}),
\end{equation}
with $E_{n,q}$ in units of $E_\mathrm{L}$.

As the atoms are initially prepared in the ground state of the lattice, for which $q=0$, we first target eigenstates in that subspace. When $q=0$, the parity of the quantum state in the $S$, $D$, $G$ ... bands is even ($c_\ell = c_{-\ell}$), while it is odd in the $P$, $F$, ... bands ($c_\ell = - c_{-\ell}$). The capability to adjust the relative phase between the different momentum components is therefore of utmost importance for such targets.

\begin{figure}[ht]
	\begin{center}
		\includegraphics[scale=1]{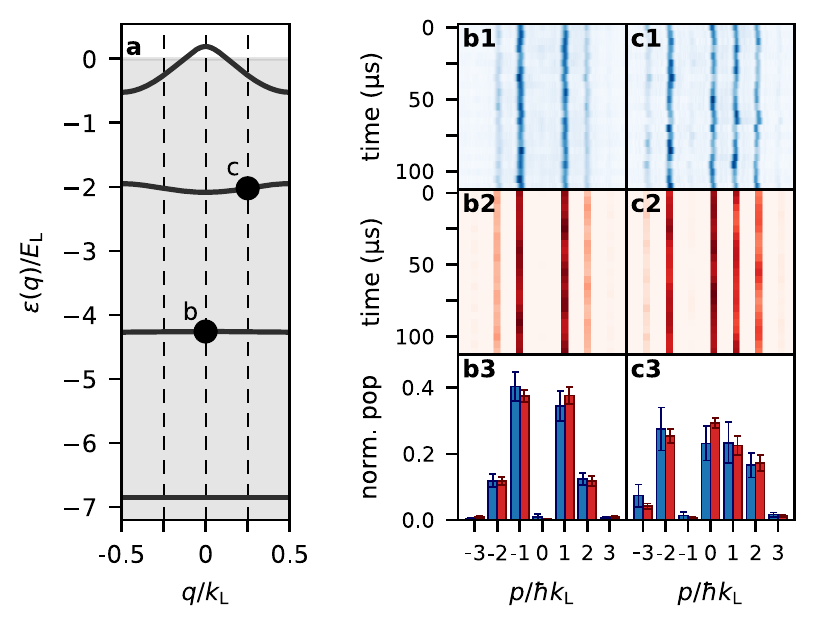}
		\caption{\textbf{Lattice eigenstates.} (\textbf{a}): Lattice band structure for a depth $s=8.2$. The grey area denotes the depth of the lattice potential. Black dots denote the states prepared in (\textbf{b}) and (\textbf{c}). (\textbf{b}): Preparation of the eigenstate of the $P$ band at quasi-momentum $q=0$. (\textbf{b1}): Experimental data showing the evolution of the prepared state in the lattice. (\textbf{b2}): Numerical evolution of the prepared state as expected from the optimal control phase. (\textbf{b3}): Time-averaged experimental (blue) and theoretical (red) momentum distributions. The error bars represent the standard deviation. (\textbf{c}): Same as (\textbf{b}) but for the eigenstate on the $D$ band at quasi-momentum $q=0.25\,k_\mathrm{L}$.}
		\label{fig:eigenstates}
	\end{center}
\end{figure}

The result of such a preparation is shown in Fig.~\ref{fig:eigenstates}\,(\textbf{b}). Here we first apply a control field $\varphi(t)$ to generate the momentum superposition corresponding to the eigenstate in the $P$ band (at $q=0$), and we monitor the evolution of the momentum distribution over the following \SI{110}{\micro\second} $\simeq 1.5\,T_0$, to reveal the lattice dynamics. The measured momentum distribution is shown in Fig.~\ref{fig:eigenstates}\,(\textbf{b1}) : as expected, since we prepared an eigenstate of the lattice potential, there is no visible evolution of the distribution. As a comparison, Fig.~\ref{fig:eigenstates}\,(\textbf{b2}) presents the evolution that one would expect from the prepared state as given numerically by the optimal control phase (which has a fidelity to the ideal $P$ band eigenstate of $\mathcal{F}_{Q,\mathrm{num}}=99\%$). Finally in Fig.~\ref{fig:eigenstates}\,(\textbf{b3}) we compare the time-averaged momentum distributions for both experimental and theoretical evolutions. We find once again good quantitative agreement between the observed and expected evolutions.

\begin{figure}[ht]
	\begin{center}
		\includegraphics[scale=1]{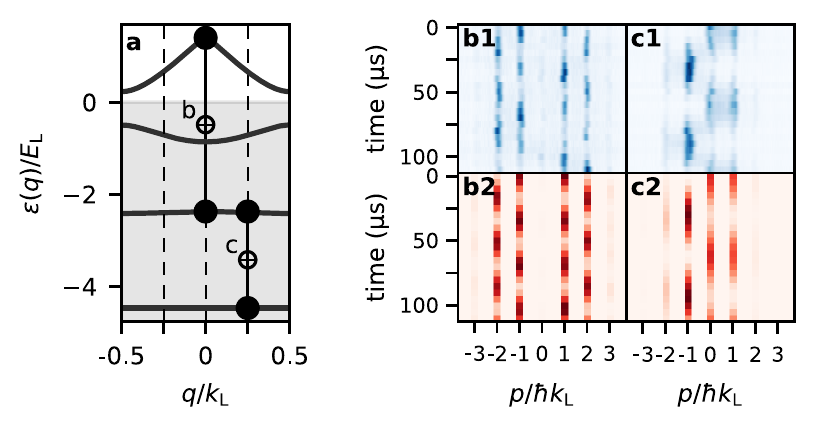}
		\caption{\textbf{Superposition of lattice eigenstates.} (\textbf{a}): Lattice band structure for a depth $s=5.6$. The grey area denotes the depth of the lattice potential. The pairs of black dots linked by a plus sign denote the states prepared in (\textbf{b}) and (\textbf{c}). (\textbf{b}):~Preparation of an equal-weight superposition of eigenstates of the $P$ and $F$ bands at quasi-momentum $q=0$. (\textbf{b1}): Experimental data showing the evolution of the prepared state in the lattice. (\textbf{b2}): Numerical evolution of the prepared state as expected from the optimal control phase. (\textbf{c}): Same as (\textbf{b}) but for an equal-weight superposition of eigenstates of the $S$ and $P$ bands at quasi-momentum $q=0.25\,k_\mathrm{L}$.}
		\label{fig:superposition}
	\end{center}
\end{figure}

We can also effectively reach lattice eigenstates at $q=q_0\neq 0$ by exploiting a change of reference frame. To that end we first use the optimal control algorithm to prepare the momentum superposition with the coefficients $c_\ell^{(n,q_0)}$ (see Eq.~\eqref{eq:bloch}). If, as in all previous experiments, the lattice was returned to a constant phase $\varphi=0$ after preparation, then the obtained state would not be an eigenstate. If however we set the lattice in motion, applying right after preparation a phase $\varphi(t>t_\mathrm{f})=2\pi (\hbar q_0/m) (t-t_\mathrm{f})/d$, we translate the prepared superposition into the $q=q_0$ subspace in the reference frame of the lattice, making it a proper eigenstate. 

In Fig.~\ref{fig:eigenstates}\,(\textbf{c}), we use this technique to prepare the eigenstate of the $D$ band at $q=0.25\,k_\mathrm{L}$. As expected, the populations of the prepared superposition do not evolve in time, demonstrating again that we have generated an eigenstate (the expected fidelity of the prepared state to the ideal eigenstate is $\mathcal{F}_{Q,\mathrm{num}}=99\%$). Note that the same state could in principle be reached by a direct optimization in the laboratory frame with an additional force breaking the translational symmetry.

Finally we can use optimal control to prepare more complex superpositions, such as the superposition of  two lattice eigenstates. This is illustrated in Fig.~\ref{fig:superposition} where we show both the experimentally measured and the calculated evolution for equal-weight superpositions of the $P$ and $F$ bands at $q=0$ (Fig.~\ref{fig:superposition}\,(\textbf{b})) and of the $S$ and $D$ bands at $q=0.25\,k_\mathrm{L}$ (Fig.~\ref{fig:superposition}\,(\textbf{c})). A striking feature of these data is the evolution of all momentum orders with a single frequency (one can contrast these with e.g. Fig.~\ref{fig:dphi}\,(\textbf{c})), which corresponds to the energy difference between the eigenstates involved.

\section{Conclusion and perspectives}

We have demonstrated a versatile optimal control scheme  for ultracold atoms in an optical lattice, that relies on the modulation of a \emph{single} parameter, the lattice position. With this simple scheme it is possible to prepare arbitrary periodic states in the lattice, with full control on the populations and relative phases of the momentum components. We have exploited this technique to reach hundreds of different targets with a high fidelity.
Interestingly, it is possible to produce in a fast manner quantum states that cannot be reached by adiabatic transformations. This method enables us to  efficiently prepare lattice eigenstates in a straightforward way, as compared to periodic modulations of phase or amplitude \cite{CabreraSpectre}, and is also robust to the presence of a small external confinement. 

This work is a promising step forward in the context of quantum simulation as it could be extended to the tailoring of the wavefunction to either load specific semi-classical orbits in the phase-space of the lattice \cite{arnal2020}, or to optimally prepare initial states for coupled atomic momentum lattices \cite{GaldwayPRL21} or lattice-based Floquet systems~\cite{Ozawa19,FloquetRMP,Nathan}. Our protocol can be readily generalized to lattices of higher dimensions. A stimulating perspective for quantum simulation is to extend this single parameter scheme to strongly interacting systems. Besides, optimal control opens new perspectives in quantum sensing, when the control function is tailored in order to magnify the response of a quantum system to a specific parameter, such as an external force.

\acknowledgments
This work was supported by Programme Investissements d'Avenir under the program ANR-11-IDEX-0002-02, reference ANR-10-LABX-0037-NEXT, and research funding Grant No.~ANR-17-CE30-0024. N.D. acknowledges support from R\'egion Occitanie and Universit\'e Toulouse III-Paul Sabatier.

\medskip

\begin{appendix}

%\section{Optimization of control time}
%\label{appen:times}
%\section{Supplemental prepared states}

\section{BEC ``printer"}
\label{appen:printer}

\begin{figure}[ht!]
	\begin{center}
		\includegraphics[width=\linewidth]{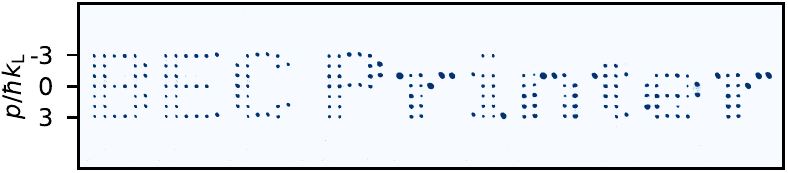}
		\caption{\textbf{Illustration of an experimental BEC dot-printer} : each of the 67 columns of the image is an absorption image taken after a QOC preparation of an equal-weight superposition of momentum states chosen among $\{\ket{\chi_i},i\in\{-3,\cdots,3\} \}$ (see text).}
		\label{fig:printer}
	\end{center}
\end{figure}

The robustness and versatility of our QOC scheme can be illustrated in an amusing and striking way through the realization of an experimental BEC ``printer". Inspired by the dot-printing technique, we can prepare and record the absorption images of all $2^7=128$ equal-weight superpositions of momenta between $p=-3\,\hbar k_\mathrm{L}$ and $p=3\,\hbar k_\mathrm{L}$. Aligning such images vertically and putting them side-by-side (in combination with an absorption image with no atoms for an empty space), we can put together letters of the alphabet, words, and sentences as we please. An example of such a printout is shown Fig.~\ref{fig:printer}.

\section{Three-momentum superpositions}
\label{appen:three}

\begin{figure}[ht!]
	\begin{center}
		\includegraphics[scale=0.9]{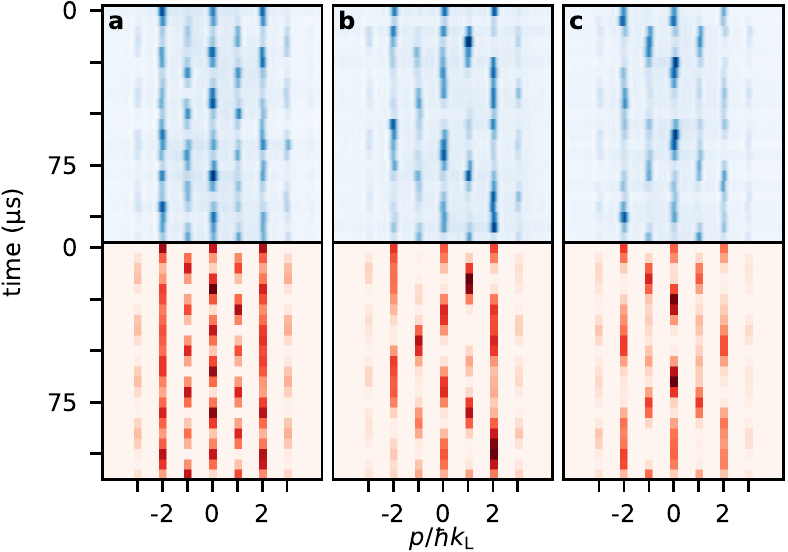}
		\caption{\textbf{Preparation of three-momentum target states.} Preparation of the superpositions  (\textbf{a}) $\ket{\psi_\mathbf{ a}}=( \ket{\chi_{-2}}+\ket{\chi_0}+\ket{\chi_2})/\sqrt{3}$, (\textbf{b}) $\ket{\psi_\mathbf{ b}} =(\ket{\chi_{-2}}+e^{2i\pi/3}\ket{\chi_0}+e^{4i \pi/3}\ket{\chi_2})/\sqrt{3}$, (\textbf{c}) $\ket{\psi_\mathbf{ c}} =(e^{2i\pi/3}\ket{\chi_{-2}}+\ket{\chi_0}+e^{2\pi/3}\ket{\chi_2})/\sqrt{3}$.  Top: stacks of integrated experimental images (blue) showing the evolution of the momentum distribution during a \SI{110}{\micro\second} holding time in a static lattice after applying the control field $\varphi(t)$ preparing the target momentum superposition. Bottom: numerical propagation  of the expected prepared state in a static lattice (red). For all three prepared states, the dimensionless lattice depth is $s=5.7\pm0.2$.}
		\label{fig:trinomes}
	\end{center}
\end{figure}

In Fig.~\ref{fig:trinomes}, we illustrate the control of the relative phases of the QOC prepared state on the following three-component momentum superpositions: $
\ket{\psi_\mathbf{ a}}= (\ket{\chi_{-2}}+\ket{\chi_0}+\ket{\chi_2})/\sqrt{3}, 
\ket{\psi_\mathbf{ b}}=(\ket{\chi_{-2}}+e^{2i\pi/3}\ket{\chi_0}+e^{4i\pi/3}\ket{\chi_2})/\sqrt{3}$ and 
$\ket{\psi_\mathbf{ c}}=(e^{2i\pi/3}\ket{\chi_{-2}}+\ket{\chi_0}+e^{2i\pi/3}\ket{\chi_2})/\sqrt{3}$.
For each of these states, we show the experimentally measured evolution of the momentum distribution in the static lattice following the state preparation (in a lattice of depth $s=5.7\pm0.2$), along with the numerically calculated evolution expected from the state as ideally prepared with QOC. The agreement between experimental and theoretical data is very good, and allows us to identify the prepared superposition.

\end{appendix}

\clearpage

\bibliography{control}

\end{document}